\documentstyle[multicol,aps,psfig]{revtex}
\draft
\def\la{\langle}
\def\ra{\rangle}
\def\beq{\begin{equation}}
\def\eeq{\end{equation}}
\def\be{\begin{eqnarray}}
\def\ee{\end{eqnarray}}

\def\k2av{\la k_T^2\ra}

\begin{document}
\title{Decisive role of fragmentation functions in hard hadron production}
\author{Xiaofei Zhang\thanks{electronic mail: xiaofei@cnr4.physics.kent.edu}
and
George Fai\thanks{electronic mail: fai@cnr4.physics.kent.edu}}
\address{Center for Nuclear Research, Department of Physics,
Kent State University \\
Kent, Ohio 44242, USA}
\author{ P{\'e}ter L{\'e}vai\thanks{electronic mail: plevai@rmki.kfki.hu}}
\address{KFKI Research Institute for Particle and Nuclear Physics,
P.O. Box 49, Budapest 1525, Hungary}
\maketitle
\vspace{-1.8in}
{\hfill {KSUCNR-202-06}}

\vspace*{1.8in}
\begin{abstract}

It is demonstrated that the fragmentation functions at large momentum
fraction play a key role in hard hadron production from relativistic
proton-proton collisions. We find that this region of the fragmentation
functions is not strongly constrained by the electron-positron 
data. This freedom can be used (together with the transverse momentum
distribution of partons) to reproduce hard pion-to-proton ratio 
data in  relativistic proton-proton collisions.

\end{abstract}
\vspace{0.2in}

\pacs{PACS Numbers: 25.75-q, 13.85.-t, 13.87.Fh}

\begin{multicols}{2}
Hadron spectra at large transverse momentum $(p_T)$, where 
perturbative Quantum Chromodynamics (pQCD) 
has good predictive power, are very important for
our understanding of the physics at the Relativistic Heavy Ion 
Collider (RHIC) and at the planned Large Hadron Collider (LHC). 
Various experiments at these colliders focus on hard (high $p_T$)
hadron spectra. Jet ``tomography'' 
(the study of the strongly-interacting medium 
via energy loss of hard partons) has been proposed 
to detect the quark-gluon plasma (QGP), using high-$p_T$ hadron 
production\cite{glv_tomo,wang_tomo}.

Suppression of total charged hadron production in $Au+Au$ collisions 
relative to a nucleon-nucleon reference is reported at 
RHIC\cite{phen_hadr,star_hadr}. However, for the 
proton-to-pion ($p/\pi$) ratio the PHENIX experiment
reports an anomalous enhancement
in $Au+Au$ collisions at $\sqrt{s}=130$~GeV\cite{phen_ppi}. 
An explanation for the $p/\pi$ enhancement was proposed recently, 
combining pQCD with soft physics and jet quenching\cite{vitev_ppi}.  
It should be kept in mind in this context that, while pQCD
is quite successful for total charged hadron ($h^+ + h^-$)
and pion production 
at large $p_T$ in $pp$ collisions, proton production in $pp$ 
is not well understood using the language of pQCD. In fact,
 pQCD underestimates the $p/\pi^+$ ratio by a factor of 3-10
in $pp$ collisions (see Fig. 3(a) of this work for $\sqrt{s}=27.4$ GeV,
and e.g. Ref. \cite{vitev_ppi} for Tevatron energy).

In this Letter we look into how pQCD is used to 
calculate $p_T$ spectra in $pp$ collisions. We 
focus on the role of a non-perturbative ingredient,
the {\it fragmentation function} (obtained by fitting 
data) in the production of hadronic final states. 
Most of the information included in the fits comes from $h^+ + h^-$ 
data. The fragmentation function (FF) of pions is  
studied in some detail. Less direct information is 
available on kaons, and the FF of protons is even less 
well-known.  In the following, we concentrate on  the proton FF 
as an example  of the role of the  FF in the hadroproduction 
of hard particles.
We find (for all types of hadrons) that 
the value of the FF in a region of phase space where it is 
least constrained plays a decisive role  
for hadron production at RHIC and LHC.

To predict the $p_T$ spectra of final state hadrons
in the framework of pQCD, perturbative partonic cross sections
need to be convoluted with parton distribution functions (PDF-s)
and fragmentation functions (FF-s) according to the factorization 
theorem\cite{Field}. 
Perturbative QCD has nothing to say about the details of FF-s,
apart from describing their scale evolution. The 
FF-s are assumed to be {\it universal}, meaning 
that once extracted from a limited set of data via a 
fitting procedure, they can then be used with
predictive power in other reactions\cite{BKK,KKP,Kretzer,BFGW}.
The situation is conceptually identical to that of the PDF-s. 
We will work with this formalism in the present paper.
As an alternative, phenomenological models (e.g. string or cluster 
models) can of course be implemented in Monte Carlo 
programs\cite{frag_mont,HIJING}.

The available fragmentation functions are obtained by fitting 
mostly $e^+e^-$ data. The FF-s extracted this way are preferred
to using information also from $pp$ ($p \bar p$) data   
to avoid the complication introduced by $k_T$ smearing
\cite{owens87,ZF_02} in the latter case. Note, however, that 
the phase space for hadronic collisions
is different from that of $e^+e^-$ reactions. 
This manifests itself in different values for $z$,
the momentum fraction of the fragmenting parton carried by the 
final hadron. In order to study the role of different regions of $z$
in the FF, we define the ratio $R_z$ as
\begin{equation}
R_z(p_T)= \int_0^z dz'{d\sigma \over p_Tdp_Tdz'}\left/
\int_0^1 dz'{d\sigma \over p_Tdp_Tdz'} \right.  \,\,\,\, ,
\end{equation}
where the cross section $d\sigma /(p_Tdp_Tdz)$ is differential in 
$p_T$ and in $z$, and the ratio depends on the upper limit of the
integration in the numerator.

\begin{figure}
\begin{center}
\psfig{figure=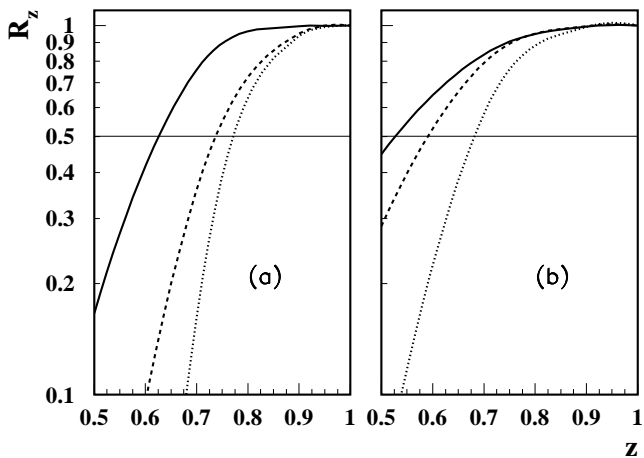,width=3.4in,height=2.5in}
\end{center}
\vspace{-0.2in}
\caption{The ratio $R_z(p_T)$ defined in Eq. (1)
as a function of $z$ for different $p_T$ values
(a) at $\sqrt{s}=27.4$ GeV with
$p_T=2$ GeV (solid line), $p_T=4$ GeV (dashed), and
$p_T=6$ GeV (dotted), and (b) at $\sqrt{s}=130$ GeV with 
$p_T=4$ GeV (solid), $p_T=8$ GeV (dashed), and
$p_T=16$ GeV (dotted).} 
\end{figure}

Figure 1 displays $R_z(p_T)$ as a function of $z$ for three different 
values of $p_T$ in the case of proton production in $pp$ 
collisions at center-of-mass (cm) energies  $\sqrt{s}=27.4$~GeV 
(part (a)) and $\sqrt{s}=130$ GeV (part (b)).
In Fig. 1(a) we show $p_T=$ 2 GeV (solid line), 4 GeV (dashed) and 
6 GeV (dotted). In Fig. 1(b), where the higher cm energy allows 
higher $p_T$ values, the solid line corresponds to $p_T=$ 4 GeV,
dashed means $p_T=$ 8 GeV, and dotted represents $p_T=$ 16 GeV.
It can be read off from Fig. 1(a) that 
when $p_T =$ 4 GeV, about 90\% of the cross section comes from the 
contribution of $z>0.6$ at $\sqrt{s}=27.4$ GeV. When $p_T =$ 6~GeV,
90\% of the cross section arises from $z>0.7$.
Fig. 1(b) shows that at $\sqrt{s}=130$ GeV, 
when $p_T =$ 4 GeV, approximately 60\% of the
cross section comes from the 
contribution of $z>0.5$; when $p_T =$ 8 (16) GeV,
70\% (95\%) of the cross section is from $z>0.5$.

In summary, the large $z$ part of the fragmentation function
dominates the proton production cross section 
in hadronic collisions.  
When $p_T$ increases, the effect becomes even more pronounced.  
This conclusion is consistent with results for
Drell-Yan processes\cite{BQZ-dy}.

The above results depend on the shape of the FF. 
If the corresponding FF-s   fall less steeply (are ``harder'')
than the KKP proton FF-s, the dominance of the large $z$  
region is even more significant.
Thus, the above effect is also very important 
for the hadroproduction of pions and kaons, which have harder FF-s.
Let us examine, therefore, what kind of 
constraints are placed on the FF-s in the large $z$ region by  
$e^{+}e^{-}$ data.

In $e^{+}e^{-}$ collisions, the hadron production cross 
section (differential in incoming momentum fraction $x$)  
is given by\cite{BKK} 
\begin{equation}
{d\sigma^h  \over dx}=\sum_{a}\int_x^1{dz\over z} 
{d{\hat\sigma}_a\over dy}(y, \mu_F, \mu_R)
D_a^h(z,\mu_F)  \,\,\,\, ,  
\end{equation}
where $d{\hat\sigma}_a/dy$ is the differential cross section of 
a partonic sub-process producing parton $a$ (as a
function of $y=x/z$, of the factorization scale $\mu_F$, and 
of the renormalization scale $\mu_R$), and $D_a^h(z,\mu_F)$
is the FF for parton $a$ to fragment into hadron $h$ with 
momentum fraction~$z$. The contributions from different partons
(quarks of flavor $i$ and gluons $g$) are summed over.

Through next-to-leading order (NLO), in the $\overline{MS}$ scheme, 
the partonic sub-process cross section for quark flavor $i$
can be written as\cite{BKK}

\be
\label{qnlo}
{d{\hat\sigma}_{q_i} \over dy}=\sigma_0 N_c {e_{q_i}}^2&&\biggl[ 
 \delta(1-y)  
+{\alpha_s(\mu_R^2)\over 2\pi} 
\bigg( P_{qq}^{0,T}(y)\ln{s\over \mu_F^2}\nonumber \\  
&&   +K_q^T(y)+K_q^L(y) \biggl)    
\biggl]    \,\,\, ,
\ee
where $\sigma_0$ is the corresponding total cross section for 
$e^{+}e^{-} \rightarrow \gamma^*\rightarrow \mu^+  \mu^-$, 
$N_c$ is the number of colors,
$e_{q_i}$ is the charge of the quark,
$\alpha_s$ is the strong coupling constant, 
and the interested reader is referred to \cite{BKK} for 
the functions $P$ and $K$. 
The corresponding expression including the 
$Z^0$ contribution is similar, but more 
complicated due to the weak coupling  of the $Z^0$\cite{epem-z}.

For gluons,
\be
\label{gnlo}
{d{\hat\sigma}_g \over dy}=\sigma_0 N_c &&\sum_{i=1}^2 N_f e_{q_{i}}^2
{\alpha_s(\mu_R^2)\over 2\pi}
\times\biggl(P_{qg}^{0,T}(y)\ln{s\over \mu_F^2} \nonumber \\
&&  +K_g^T(y)+K_g^L(y) \biggl)  \,\,\,\, ,
\ee
in a similar notation ($N_f$ is the number of active flavors). 

The usual parameterization of the fragmentation functions
at an input scale $\mu_0$ is\cite{BKK,KKP,Kretzer,BFGW}
\begin{equation}
D(z,\mu_0^2)=Nz^\alpha(1-z)^\beta   \,\,\,\,   ,
\label{Dpar}
\end{equation}
where $\alpha$ and $\beta$ are fixed by fitting to a given set of data.
The parameter $\alpha$ describes the behavior of the FF in the small 
$z$ region, while $\beta$ determines the behavior in the large $z$ region. 
The FF-s are not as well studied as the PDF-s. The 
KKP fragmentation functions\cite{KKP} provide
one of the few sets that contain 
FF-s for protons. Since we are particularly  
interested in proton production, we will use KKP FF-s\cite{KKP} in the
following calculations. Figure 2 illustrates how arbitrary changes in the  
value of $\beta$ modify the proton FF-s in $e^{+}e^{-}$ collisions
in the large $z$ region, and how these modifications 
influence the agreement with the experimental data
at $\sqrt{s}=91$ GeV\cite{sld}.

In the leading order approximation, one has $x=z$.
Therefore, at sufficiently high cm energies, $e^{+}e^{-}$ 
data points at large $x$ should constrain the fragmentation function
at large $z$. It can be seen from Fig. 2 that the number of available
data points at large $x$ is quite limited, 
and that the uncertainty associated with the highest-$x$ point 
in particular is rather large. This is not surprising, since 
this point is close to the phase space limit of the experiment.
Since there is no reason to expect that a NLO
calculation would change the state of affairs at  the
high energies considered here,
we use the leading order pQCD results in the following discussion.
We use  $\mu_F=\sqrt{s}$ (this sets the log terms in (\ref{qnlo})
and (\ref{gnlo}) to zero).

\begin{figure}
\begin{center}
\psfig{figure=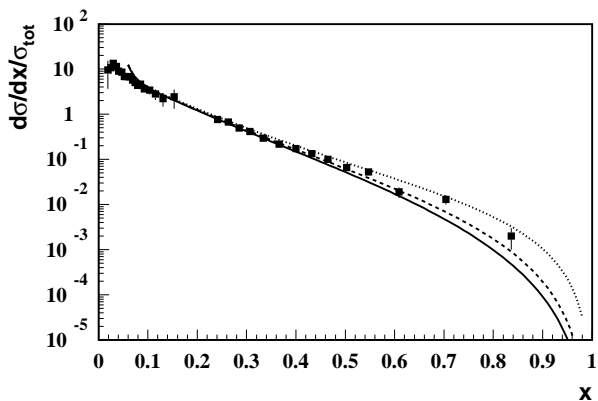,width=3.3in}
\end{center}
\vspace{-0.3in}
\caption{Perturbative QCD calculations using FF-s with 
different large $z$ behavior compared to $e^+e^-$ data at 
$\sqrt{s}=91$ GeV \protect\cite{sld}. The solid line is the result
with $\beta=\beta_{KKP}$ \protect\cite{KKP}.
The dashed (dotted) line represents $\beta=0.8\beta_{KKP}$
($\beta=0.5\beta_{KKP}$) for modified $u,d$ ($\bar u,\bar d$) quark
and gluon FF-s.}
\end{figure}

Figure 2 illustrates that there is significant freedom in the choice
of the large $z$ behavior of FF-s based on  $e^{+}e^{-}$ experiments. 
To clearly display the point that these data     
can  be fitted by FF-s 
being very different in the large $z$ region,
we show the cross section calculated with the KKP value\cite{KKP}
of $\beta$ in Eq. (\ref{Dpar}) (solid line), together with
results with $\beta=0.8\beta_{KKP}$ (dashed) and
$\beta=0.5\beta_{KKP}$. We conclude from Fig. 2 that 
taking  $\beta=0.8\beta_{KKP}$ does not change 
the quality of the fit to the $e^+e^-$  data at this energy.
It is also clear from the above discussion, 
how different parameterizations of FF-s\cite{BKK,KKP,Kretzer,BFGW} 
can be rather different
in the large $z$ region even for $h^+ + h^-$ fragmentation 
(the FF-s studied in most detail).

The situation concerning gluon FF-s at large $z$ 
is even less certain than it is  
for quarks. The contribution of gluon
fragmentation to the $e^{+}e^{-}$ hadron production 
cross section appears only as a NLO correction.
In hadronic collision, the contribution from gluon processes
is much more important than in $e^{+}e^{-}$\cite{BQZ-dy}.
Since the probability of finding a gluon in the proton
(the gluon  PDF) increases rapidly as $ x\sim 2p_T/\sqrt{s}$
decreases, gluon fragmentation plays an amplified role
at RHIC and LHC energies, compared to fixed target  energies.
It is thus very important to note that  
the large $z$ gluon FF for $h^+ + h^-$ 
obtained including some $pp$ data can be an order of magnitude 
larger than the one not including $pp$ information\cite{BFGW}.

As we have seen, the $e^{+}e^{-}$ data do not strongly constrain the 
FF-s in the large $z$ region. It is therefore felt that one has
some freedom to fit the proton $p_T$ spectra in $pp$ collisions by 
modifying the large $z$ behavior of the proton fragmentation functions.
To illustrate this idea,
we compare the results for the $p/\pi^+$ ratio calculated using 
the KKP form of the proton FF-s, but varying the value of
the parameter $\beta$. In addition, the transverse momentum 
distribution of the partons in the proton needs to be taken into
account in a more complete calculation. As long as no 
first-principles treatment of this effect is available, the
width of the transverse-momentum distribution $\k2av$
provides another phenomenologically adjustable parameter.

\begin{figure}
\begin{center}
\psfig{figure=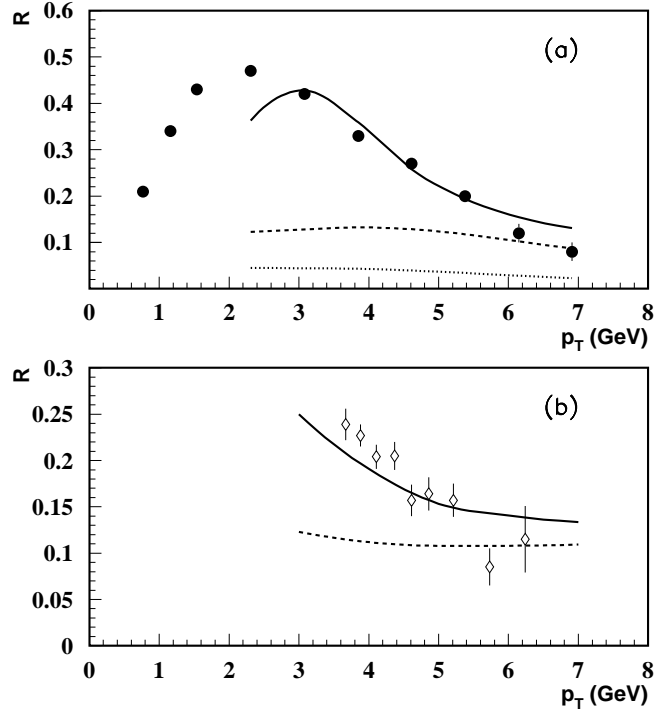,width=3.5in}
\end{center}
\vspace{-0.3in}
\caption{Comparison of $p/\pi$ ratio data and pQCD results in $pp$ 
collisions: (a) at $\sqrt {s}=27.4$ GeV ( dotted line is 
$\beta=\beta_{KKP}$, dashed is $\beta=0.8\beta_{KKP}$, 
 solid line is calculated with different $\k2av$ values
for the pion and the proton as described in the text; 
 data are from Ref. \protect\cite{antr-79}),
(b) at $\sqrt {s}=61$ GeV (ISR data \protect\cite{isr}) 
and prediction for RHIC at $\sqrt{s}=130$ GeV (dashed line).}
\end{figure}

Figure 3 demonstrates that the remaining pQCD freedom represented by 
the above parameters may be sufficient to achieve a better description 
of the available data on $p/\pi^{+}$ ratios, without invoking any 
other mechanism. Our goal here is not to fit the data; Fig. 3 serves
only as an illustration. The dotted line in  
Fig. 3(a) shows that the calculated $R=p/\pi^{+}$ ratio 
using the original KKP proton FF-s underestimates the data (dots)
\cite{antr-79} 
at $\sqrt{s}=$ 27.4 GeV by up to a factor of 10. 
However, if we set $\beta=0.8\beta_{KKP}$ in the proton FF (dashed line),
the pQCD result for the $p/\pi^+$  ratio comes    
close to  the experimental data for $p_T>6 $ GeV. 
In the above calculation, we used $\k2av=0.9$ GeV$^2$
for both proton and pion production. 
However, it is not clear that the input $\k2av$ for proton 
and pion production needs to be identical. Due to the
large mass of the proton, the effective $\hat s$ (the
energy involved in the partonic cross section) is larger for proton 
production than for pion production at the same $p_T$.
Large $\hat s$ leads to more room for a dynamical intrinsic $k_T$,
just like in the Drell-Yan case, where the larger $Q^2$ of the lepton
pair leads to a larger $k_T^2$. 
Therefore it is conceivable that a larger input $\k2av$ is needed 
for proton production than for pion production\cite{QZ_resum}. 

Using $\k2av_p=2.6$ GeV$^2$ for the proton and keeping the 
value $\k2av_{\pi}=0.9$ GeV$^2$ for the pion in the calculation,
we obtain the solid line in Fig. 3(a) for the $p/\pi^+$  ratio
in $pp$ collisions. It is fair to say that the result of this 
calculation is in satisfactory agreement with the data for
$p_T \gtrsim 3$ GeV. (Below $p_T \approx$ 2-3 GeV pQCD is not 
considered reliable, and non-perturbative effects may become
important\cite{junction}.) We obtain similar
results comparing to data at $\sqrt {s}=38.8$ GeV. 

In Fig. 3(b) we explore the energy dependence of the above 
proposition. Here, the solid line is the result of a
pQCD calculation for the $p/\pi^+$  ratio
in $pp$ collisions at $\sqrt{s}=61$ GeV,
with $\beta=0.8\beta_{KKP}$ and $\k2av_p=3$ GeV$^2$ 
(while keeping $\k2av_{\pi}$ constant). The calculated results
are compared to ISR data (open symbols)\cite{isr}.
We also give the prediction for RHIC energy, $\sqrt{s}=$ 130 GeV,
with $\k2av_p=$ 3 GeV$^2$ (dashed line).

The main purpose of this paper is to call attention to  the importance 
of the behavior of fragmentation functions in the large $z$ region
 for hard hadron production.
Recently, similar ideas were considered for 
bottom hadroproduction at the Tevatron in terms of the
moments of the fragmentation function (higher moments emphasizing
the importance of large $z$)\cite{bottom}. 

In conclusion, since the large $z$ details of the fragmentation functions 
are very important for pQCD predictions of high $p_T$
hadron production in both $pp$ and $AA$ collisions at RHIC and LHC,
a comprehensive study of these fragmentation functions with particular
attention to the large z region is strongly warranted. Additional
complications not addressed in the present note include e.g. the
effect of changing $\beta$ on the scale dependence, a first-principles
treatment of the transverse momentum degree of freedom, and various nuclear
effects. These are left for future work. It appears that available
$p/\pi$ ratios from $pp$ collisions at high $p_T$ can be reproduced by 
adjusting the large z behavior of the fragmentation functions and the width of 
transverse momentum distributions.  

We are grateful to M. Gyulassy, S. Kretzer,  
and J. W. Qiu 
for stimulating discussions. 
This work
was supported in part by the U.S. DOE under DE-FG02-86ER-40251, by
the NSF under INT-0000211, and by Hungarian OTKA under T025579 and
T032796.

\end{multicols}

\begin{references}
\bibitem{glv_tomo}
M. Gyulassy, P. Levai, I. Vitev, nucl-th/0112071 

\bibitem{wang_tomo}
E. Wang and  X.-N.  Wang,  hep-ph/0202105. 

\bibitem{phen_hadr}
K. Adcox {\it et al.}, Phys. Rev. Lett. {\bf 88},
 022301 (2002).

\bibitem{star_hadr}
J. Harris, Nucl. Phys. {A698}, 64c (2002).

\bibitem{phen_ppi}
K. Adcox {\it et al.}, nucl-ex/0112006.

\bibitem{vitev_ppi}
I. Vitev and M. Gyulassy, Phys. Rev.  C {\bf 65}, 041902 (2002).

\bibitem{Field}
R.D. Field, {\it Applications of Perturbative QCD}, Addison-Wesley, 1995.





\bibitem{BKK}
J. Binnewies, B.A. Kniehl, and G. Kramer, Z. Phys. C {\bf 65}, 471 (1995).

\bibitem{KKP}
B.A. Kniehl, G. Kramer, and B. P{\"o}tter, Nucl. Phys. B {\bf 597}, 337 (2001).

\bibitem{Kretzer}
S. Kretzer, E. Leader, and E. Christova,  Eur. Phys. J. C {\bf 22}, 
 269, (2001). 

\bibitem{BFGW}
 L. Bourhis, M. Fontannaz, J.P. Guillet, and M. Werlen 
Eur.Phys.J.C {\bf 19},  89  (2001). 

\bibitem{frag_mont}
G. Marchesini {\it et al.}, Comp. Phys. {\bf 67}, 465 (1992);
T. Sjostrand, Comp. Phys. {\bf 82}, 74 (1994);
S. Chun and C. Buchanan, Phys. Rep. {\bf 292}, 239 (1998). 

\bibitem{HIJING}
X. -N. Wang and M. Gyulassy, Phys. Rev. Lett. {\bf 68}, 1480 (1992).

\bibitem{owens87}
J.F. Owens, Rev. Mod. Phys. {\bf 59}, 465 (1987).

\bibitem{ZF_02}
Y. Zhang, G. Fai, G. Papp, G.G. Barnafoldi, and P. Levai, Phys. Rev. C
{\bf 65}, 034903 (2002).
\bibitem{BQZ-dy}
E. Berger, J. W. Qiu, and X. F. Zhang,  Phys. Rev. {\bf D65}, 034006  (2002). 

\bibitem{epem-z}
P. J. Rijken and W. L. van Neerven, Nucl. Phys. {\bf B487}, 233 (1997).


\bibitem{sld}
K. Abe, {\it et al.}, (SLD collaboration), Phys. Rev. {\bf D59}, 052001
(1999). 

\bibitem{antr-79} 
D.Antreasyan, {\it et al.}, Phys. Rev. {\bf D19}, 764 (1979).
 




\bibitem{QZ_resum}
J.W. Qiu and X.F. Zhang, Phys. Rev. Lett. {\bf 86}, 2724 (2001);
Phys. Rev. D {\bf 63}, 114011 (2001).

\bibitem{junction}
C. Adler {et al.}, Phys. Rev. Lett. {\bf 87}, 112303 (2001);
G. C. Rossi and G. Veneziano, Phys. Rep. {\bf 63}, 153 (1980);
D. Kharzeev, Phys. Lett. {\bf B 378}, 238 (1996).  


\bibitem{isr}
A. Breakstone {\it et al.}, Z. Phys. C {\bf 36}, 567 (1987).
\bibitem{bottom}
M. Cacciari and P Nason, hep-ph/0204025.

\end{references}
\end{document}